# Substrate-Selective Adhesion of Metal Nanoparticles to Graphene Devices

Patrick J. Edwards[1,2], Sean Stuart[2], James T. Farmer[1], Ran Shi,[3]
Run Long,[3] Oleg V. Prezhdo[1,4], Vitaly V. Kresin[1]

[1] Department of Physics and Astronomy, University of Southern California, Los Angeles, CA 90089-0484, USA

[2] Physical Sciences Laboratories, The Aerospace Corporation, 355 S. Douglas St., El Segundo, CA 90245, USA

[3] College of Chemistry, Key Laboratory of Theoretical and Computational Photochemistry of Ministry of Education, Beijing Normal University, Beijing 100875, China

[4] Department of Chemistry, University of Southern California, Los Angeles, CA 90089, USA

## Abstract

Nanostructured electronic devices, such as those based on graphene, are typically grown on top of the insulator $SiO_2$. Their exposure to a flux of small size-selected silver nanoparticles has revealed remarkably selective adhesion: the graphene channel can be made fully metallized while the insulating substrate remains coverage-free. This conspicuous contrast derives from the low binding energy between the metal nanoparticles and a contaminant-free passivated silica surface. In addition to providing physical insight into nanoparticle adhesion, this effect may be of value in applications involving deposition of metallic layers on device working surfaces: it eliminates the need for masking the insulating region and the associated extensive and potentially deleterious pre- and postprocessing.

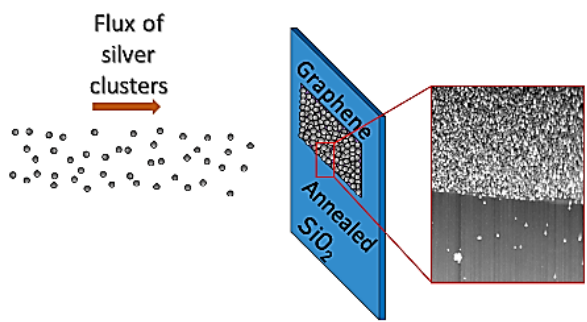

Graphene has been extensively explored as a material in microelectronic devices, particularly in the field effect transistor (FET) architecture. Its gapless semimetal nature enables fabrication of devices exhibiting ambipolar conduction, with high carrier mobilities tunable via electrostatic gating. Additionally, the electronic structure of the devices' exposed conducting graphene channel has been shown, for better or worse, to be easily affected by surface adsorbates. This fact has led to many studies aiming to use adsorbates to modify the carrier populations or bandgap of graphene FETs in order to adapt these devices to various applications in a controlled way (for examples, see refs[1–6]).

For the purpose of device modification, metallic nanoparticles and nanoclusters display many properties that make them suitable as graphene surface dopants. Their electron affinities and ionization energies are material- and size-dependent, making it possible to vary the amount of charge accepted from or donated to the graphene device.[7,8] Nanoparticles possess strong optical resonances, enabling photoinduced charge transfer into the device[9,10] and a range of other plasmonics-based graphene device applications[11–13] such as photocatalysis, solar energy conversion, photodetection, and surface-enhanced Raman scattering. Furthermore, they can serve as anchors for external adsorbates, enhancing the devices' functionality as sensors.[14–17] Another interesting application is the capability of the graphene FET channel to detect the superconducting transition at the nanoscale.[18–20]

Bare metal nanoparticles deposited from the gas phase represent a particularly promising category of dopants. In contrast to their colloidally derived counterparts, gas phase particles do not require surface ligands used to prevent coalescence in solution. As such, there is no influence or contamination from ligands or their remnants, and no need for harsh post-processing required for their removal. Furthermore, there exist methods (such as the magnetron sputtering/condensation source utilized in the work described below) of producing directed beams of nanoparticles of a variety of materials, comprising either single elements or alloys with controllable species ratios.[21–23] The nanoparticles in the beam are typically electrically charged, making it possible to filter them by size using mass spectrometry, measure their flux for accurate dosing, and adjust their kinetic energy for surface soft-landing (or for energetic implantation, if desired).

However, there is a drawback to decorating graphene channels by using a flux of nanoparticles, especially if multiple devices are fabricated on a single oxide wafer: upon exposure to the beam, the entire exposed surface, including the insulating region, may become metallized by a nanoparticle film. The conventional solution would employ lithographic techniques, in which the devices are covered with a resist, a mask is developed in it so as to expose only the devices' graphene channels, and the resist is removed after nanoparticle deposition. Unfortunately, lithographic resists used to pattern devices have been shown to alter the electronic properties of graphene and have proven to be very difficult, if not impossible, to fully remove once introduced. Furthermore, the available methods, such as acid treatments[24] and high temperature baking,[25] risk damaging the dopant nanoparticles, substantially and uncontrollably altering their properties and their effect on the device.

In this work we describe beam deposition of silver nanoparticles onto graphene on a silicon dioxide substrate. We discovered that when the devices are carefully cleaned prior to deposition, the particles almost exclusively coat the graphene and not the surrounding $SiO_2$. As a consequence, graphene devices can be decorated and doped by metal nanoparticles and nanoclusters of variable sizes without the need of *any* post-fabrication lithographic patterning.

Devices were fabricated from a commercially produced wafer of graphene grown by chemical vapor deposition (CVD) on a Si/SiO$_2$ substrate with a 285 nm oxide layer (Grolltex). A rectangular channel was etched out of the graphene layer, uncovering the surrounding silica surface. Details of the procedure and an image of the device can be found in the Supporting Information. Prior to nanoparticle deposition, the devices baked at 500° C for 24–36 hours in a $10^{-10}$-$10^{-9}$ mbar vacuum chamber to remove residues of EBL resist (polymethyl methacrylate resin, PMMA) and other possible contaminants from surface.[25]

For trials that involved only blank silica surfaces, Si/SiO$_2$ wafers (MTI Corp.) were diced and cleaned via successive 30 minute ultrasonications in acetone (twice), methanol (twice), isopropyl alcohol (twice), and deionized water. During sonication, these blocks ("dies") were held in covered test tubes to prevent any particulates from collecting on the liquid meniscus and potentially transferring to the dies upon removal. As will be demonstrated, although acetone treatment is considered standard for PMMA removal[26–28] even this thorough procedure leaves some "sticky islands" on the surface which affect its adsorption properties.

A nanoparticle beam is produced using a DC magnetron gas aggregation source.[22,29] This system consists of three segments: a gas-aggregation source chamber, a mass filter, and a deposition chamber. The base pressure in all chambers prior to deposition was approximately $10^{-9}$ mbar. In the source (Mantis Nanogen), a silver target (99.99%, ACI Alloys) is sputtered by an argon plasma at a power of 15–25 W. The resulting metal vapor condenses into nanoparticles while transported inside a liquid nitrogen-cooled aggregation region by a flow of inert argon and helium gas. The data presented below were compared for various ratios of the flow rates of Ar/He through the source (140/70, 140/140, 70/70, 70/140 and 70/10, all values in sccm), and the conclusions were independent of this ratio. During operation, because of this gas flow the source and deposition pressures rose to ~$10^{-4}$ mbar and ~$10^{-5}$ mbar, respectively. Thanks to the configuration of the magnetron block, the majority of nanoparticles are negatively charged.[30,31]

Upon exiting the condensation source, the resulting beam passes through a quadrupole mass filter (Mantis MesoQ) and enters the deposition chamber where samples, mounted on a linear translation stage, can be exposed to it one at a time. The beam is collimated by an aperture, and the nanoparticle ion flux is determined with the help of a picoammeter (Keithley 6487). This establishes the exposure time required for a desired coverage.

After deposition, surface imaging was conducted via atomic force microscopy (AFM, Asylum Research Cypher ES). In order to avoid disturbing the weakly bound nanoparticles, the cantilever drive amplitude and setpoint were tuned to the regime of imaging using long-range attractive forces.[32] Scanning electron microscopy (SEM, FEI Nova NanoSEM 450) verified the results, as described below.

To gain insight into the expected coverage of nanoparticles on the graphene FETs, initial depositions were performed on separate small wafers of baked CVD graphene and of cleaned Si/SiO$_2$ which were mounted on the same sample puck and exposed to the beam simultaneously. The mass spectrometer was set to select Ag nanoparticles of 6–8 nm diameter (~7–15×10$^3$ atoms) for deposition, at nominal coverages ranging from 0.05 to 0.2 of a monolayer. The estimated 140 m/s velocity of the beam[33] implies a deposition energy of ~10 meV per atom. To rule out any possible role of beam profile inhomogeneity, some of the samples also were prepared with the wafers interchanged.

In every single trial, the observed density of nanoparticles on graphene significantly exceeded that on SiO$_2$. even though they were exposed to the same beam flux. Figure 1 contains representative comparison images of SiO$_2$ (a) and graphene (b) surfaces.

To confirm the AFM scan settings and rule out any potential bias due to tip-particle interaction, these observations were verified by SEM imaging. Figure 2a reveals that nanoparticles appear only in those areas of the SiO$_2$ substrate where some surface contamination exists, plainly visible as a film-like constrast around the dark nanoparticle dots. As mentioned above, these patches represent post-lithography or post-solvent residue which outlasted the cleaning process. On the other hand, clean SiO$_2$ regions contain no adsorbed nanoparticles.

The adjoining graphene samples were also examined (Figure 2b), confirming significantly greater nanoparticle coverage in every sample. Additionally, in this figure one finds occasional gaps in the graphene (crack defects or empty pits) which expose the underlying SiO$_2$. These patches, which are residue-free, again show no sign of nanoparticle attachment.

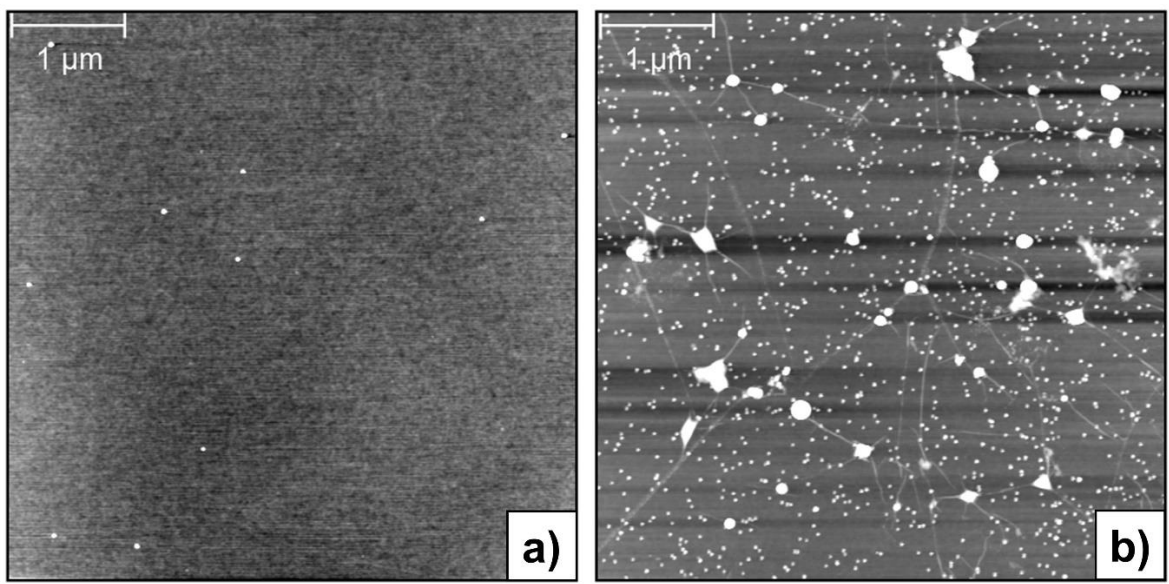

**Figure 1.** AFM imaging comparison of 8 nm Ag nanoparticle deposition on SiO$_2$ (a) and graphene (b) under the same deposition conditions. The scan width of all images is the same, to illustrate the manyfold increase in nanoparticle density adsorbed on the graphene samples. (The larger white blotches on the graphene surface are PMMA patches which collected at crack defects, common to CVD graphene, and did not fully desorb during the baking process.)

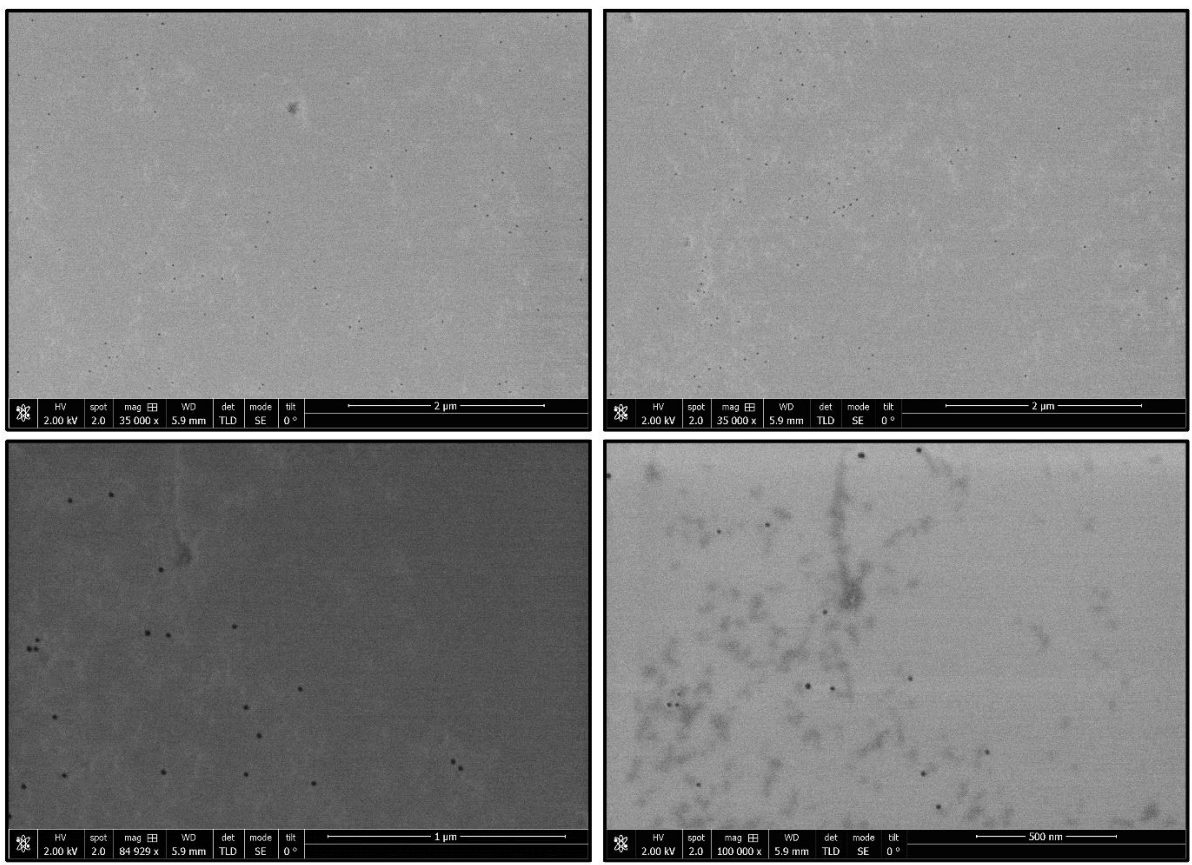

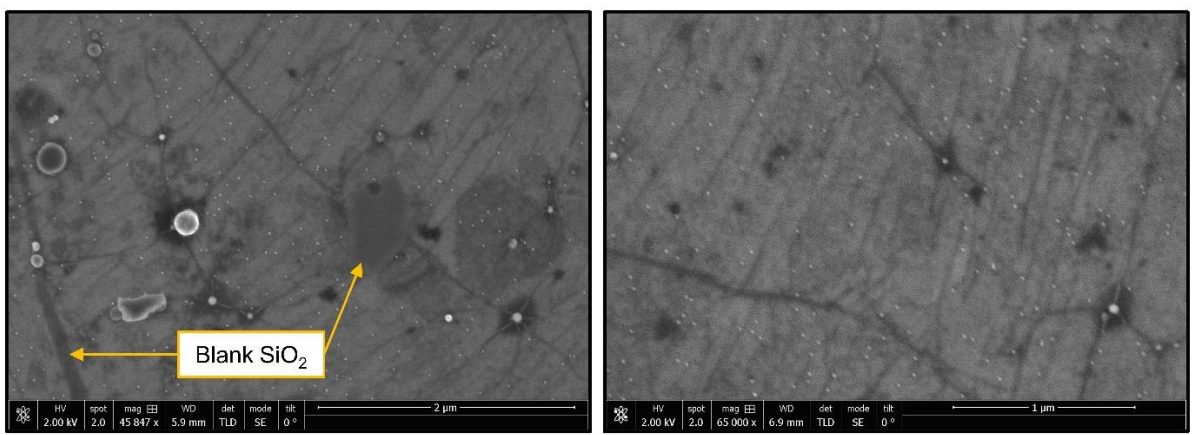

**Figure 2.** a) SEM characterization of SiO$_2$ surfaces after Ag nanoparticle deposition, confirming that the particle adsorption is low and documenting that it is confined to areas containing splotches of surface lithography residue. b) SEM characterization of graphene surface after 0.1 monolayer Ag nanoparticle deposition. Ag clusters are observed on the graphene surface as small white dots and a complete absence of deposition is noted along the exposed SiO$_2$ underlayer.

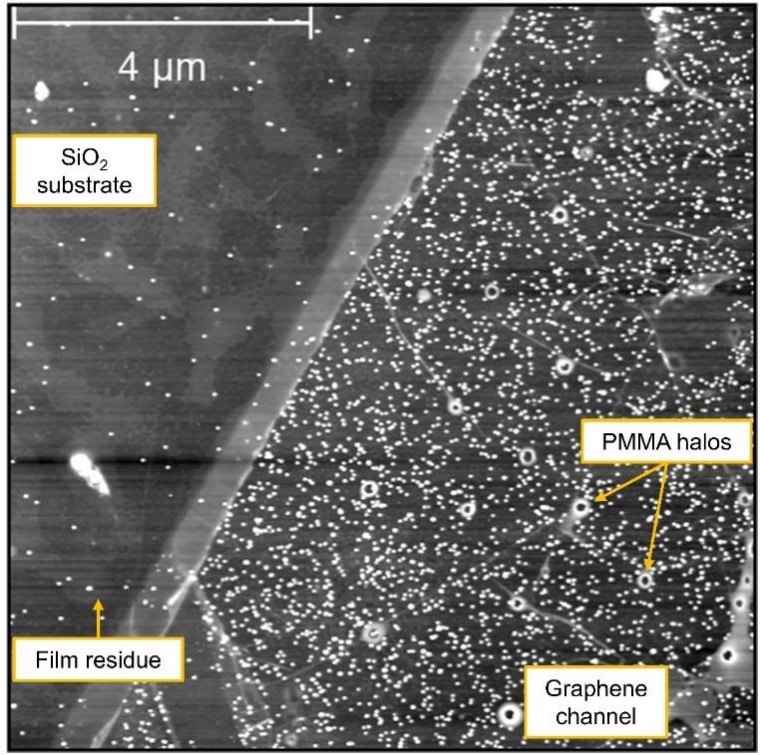

**Figure 3.** AFM imaging of dilute Ag nanoparticle deposition near the graphene channel boundary in the FET device. Again a large population difference is observed between the two substrates. Without heat treatment, islands of PMMA residue persist on the surface, and once again nanoparticles on $SiO_2$ are observed only on them and not on pristine $SiO_2$.

Figure 3 shows the result of nominally 20% coverage of 10-20 nm silver nanoparticle onto a graphene FET device without a pre-baking treatment. Consistent with the previous results, AFM imaging reveals a very large contrast between the nanoparticle populations on $SiO_2$ and graphene areas of the sample, with the former population significantly lower than might have been anticipated and restricted to areas with residue contamination.

Finally, we explored whether the effect remains robust for heavier coatings, which would otherwise lead to complete metallization of the surface. We deposited 6–8 nm Ag nanoparticles onto graphene devices at nominal coverages of 3 and 6 monolayers. For comparison, in each deposition cycle one device was subjected to high vacuum bake-out prior to deposition, while another was left "as fabricated." Figure 4a summarizes the results for 6 monolayer deposition, which is qualitatively the same as the 3 monolayer data shown in Supporting Figure S2. The image clearly displays a stark contrast in surface coverage as a consequence of pre-treatment. In the "as processed" device the Ag deposition is continuous throughout, nearly to the point of obscuring the graphene step edge. In contrast, the pre-cleaned device shows a heavy coverage of Ag on the graphene channel but few nanoparticles on the surrounding oxide. This figure underscores the main results of this work.

The images in Figures 2-4a uniformly point to the fact that incoming nanoparticles are disinclined to settle on contaminant-free $SiO_2$ surfaces.

Because most of the nanoparticles arrive as ions, one may inquire whether the effect may be due to electrostatic charging of the oxide surface. The following observations verify that this is not the case. First, as was shown in Figure 2b even patches of clean $SiO_2$ surrounded by graphene remain free of nanoparticles despite being too small to set up a strong electric field by themselves. Second, we imaged an "inverse" configuration which had small and electrically isolated patches of graphene surrounded by $SiO_2$, and found that the former became fully covered by nanoparticles whereas the latter did not (see Figure 4b and the Supporting Information). If the oxide had been exerting a strong electrostatic repulsion, it would have prevented the nanoparticles from reaching the graphene. These images represent another striking illustration of the contrast in nanoparticle adsorptivity between $SiO_2$ and graphene. Third, we performed an analogous deposition experiment using mica, which also is an insulating surface, and found that in contrast with silica it becomes fully coated with nanoparticles (see Supporting Figure S4).

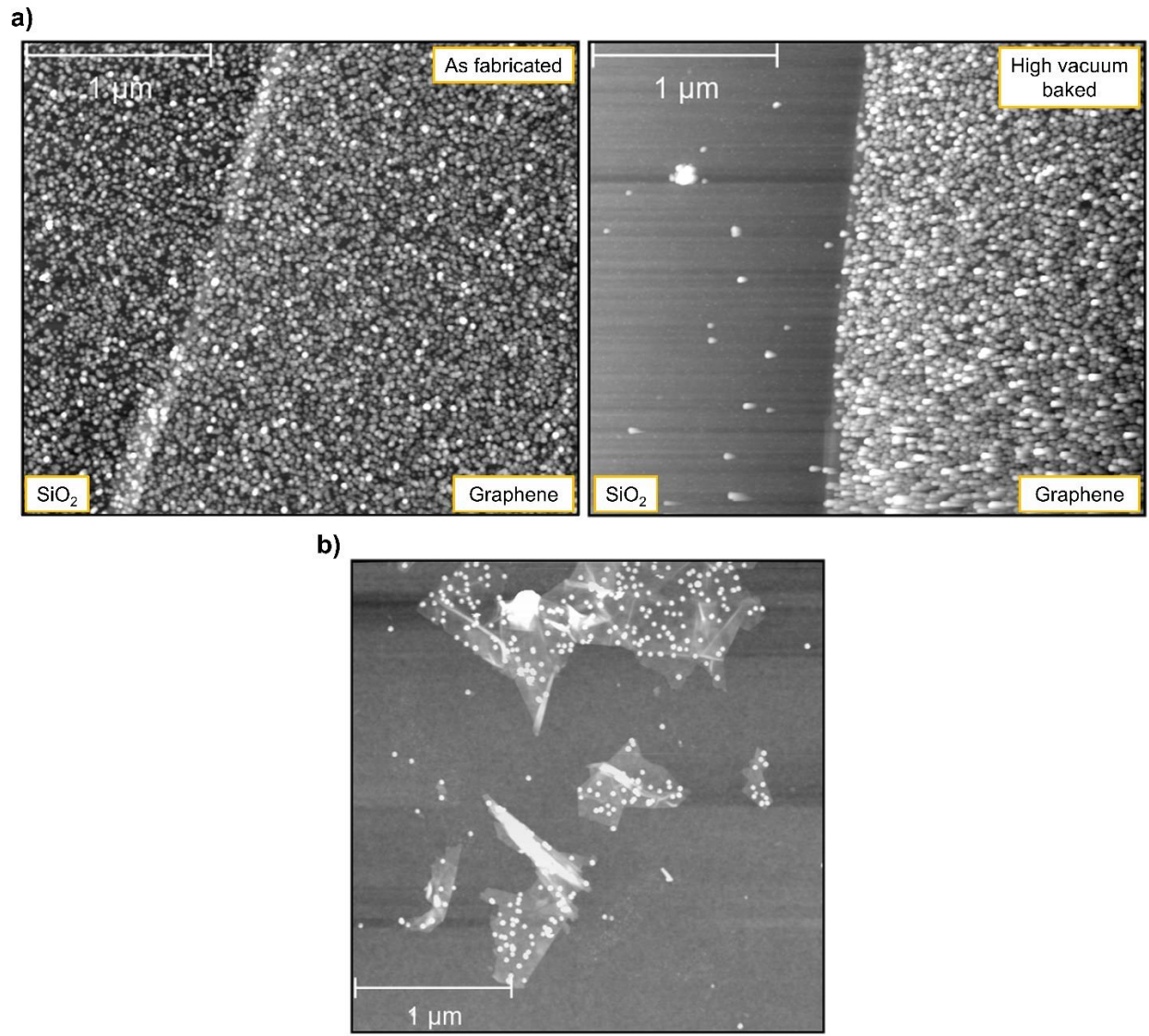

**Figure 4.** a) AFM images of graphene FET devices after the deposition of 6 monolayers of Ag nanoparticles. An extremely high level of adhesion selectivity is observed for residue-free surfaces. b) Nanoparticles are covering electrically isolated pieces of graphene but not the surrounding expanse of $SiO_2$.

A purely diffusion-based mechanism for the lack of surface adsorption on clean $SiO_2$ surface also can be ruled out. Without any graphene to diffuse toward, one would expect to see some form of deposition on the oxide wafers. With graphene present, diffusion would lead a pile-up of nanoparticles along the edge boundaries,[31,34,35] which is not observed in any images (see an additional illustration in Supporting Figure S3).

We can therefore surmise that the behavior of silver nanoparticles on clean $SiO_2$ surfaces derives from very weak attraction between the two.

This conclusion was further supported by the observation that such nanoparticles were easily lifted off when the AFM was switched to contact-mode nanomanipulation. As described in the Supporting Information, in this mode we observed nanoparticles desorbing from pre-treated silica surfaces in favor of adhering to an iridium-coated tip, something not seen upon deposition onto unbaked devices.

Taken as a whole, the results presented above point to the treatment of the $SiO_2$ surface prior to deposition as the dominant factor in adsorption. They suggest that clean $SiO_2$ is inherently uninterested in attaching silver nanoparticles.

Poor adhesion, though not in this extreme, has been observed in atomic deposition of silver thin films onto oxide surfaces.[36–38] It is common practical knowledge in the device fabrication industry that when depositing Ag or Au electrical contact pads onto silicon oxide wafers it is necessary first to deposit a thin intermediary "adhesion layer" such as Cr or Ti.[36] Without it, noble atom films degrade in ambient conditions over time and eventually delaminate from the surface.[37] It has been proposed to make use of this phenomenon for fabricating multilevel interconnects in semiconductor devices.[39]

While they are impermanent, atomic layers do coat silica surfaces at least temporarily. This is in contrast to the present nanoparticle deposition data, therefore size effects also must play a role. This is consistent with the behavior of atomic noble-metal films: with time these films coarsen and form metal domains which then separate from the $SiO_2$ surface.[37] Thus the phenomenon reported in our work involves an interplay between an inherently weak attractive force and finite-size effects. These factors have been explored by model calculations described below.

In order to elucidate the atomistic origin of the observed effect within a practicable calculation, we compute the energy of interfacial binding between a silver nanoparticle and graphene, and between the nanoparticle and $SiO_2$ surfaces of different roughness. The binding energy

$$E_{bind} = (E_{Ag} + E_{slab}) - E_{Ag/slab}$$

is calculated by subtracting the energy of the interacting system ($E_{Ag/slab}$) from the sum of the energies of the bare slab ($E_{slab}$), *i.e.*, graphene or $SiO_2$, and the nanoparticle. A positive binding energy reflects stronger chemisorption following the usual convention.[40,41]

To estimate the strength of the interfacial interaction, we select a pyramidal $Ag_{20}$ nanocluster, which represents a typical model system,[42–45] interfaced with either a 6×6 (001) graphene sheet or a 3×4 (001) $SiO_2$ surface. Details of the simulation are presented in the Supporting Information. Long-range forces, which may become more relevant for the larger nanoparticles, are not fully treated by the simulation, but its general conclusions are expected to be applicable.

The undercoordinated Si and O atoms are passivated with –OH and –H, respectively, given that the (0 0 1) surface of the α-quartz $SiO_2$ crystalline structure is hydrophilic[46,47] and is exposed to water vapor during substrate mounting. (Even if the sample remained in the high vacuum chamber, water molecules would be present in sufficient quantities to promptly saturate the surface bonds.) This aspect is important because passivation reduces the particle-substrate bonding energy by a factor of 5-10.

Using the optimized geometries shown in Figure 5, we obtain the binding energies of 3.71 eV

for Ag$_{20}$/graphene (Figure 5a), and 0.39 eV, 1.39 eV, and 0.91 eV for the Ag$_{20}$ nanoparticles interfaced with smooth (Figure 5b), slightly rough (Figure 5c) and strongly rough (Figure 5d) SiO$_2$ surfaces. The interaction is much stronger in the Ag$_{20}$/graphene system than at the Ag$_{20}$/SiO$_2$ interface, rationalizing the experimental observations that Ag nanoparticles attach strongly to the graphene. Interestingly, the binding energy is rather sensitive to the roughness of the SiO$_2$ surface. The smooth SiO$_2$ surface gives the weakest interaction, an order of magnitude weaker than graphene. The interaction is most favorable when the SiO$_2$ surface is slightly rough. Excessive roughness weakens the interaction.

The strength of the interfacial interaction correlates well with the amount of charge transfer from Ag$_{20}$ to the substrate, as characterized by the charge density difference and Bader charges.[48] The electron accumulation and depletion regions are represented by yellow and blue colors, respectively (Figure 5e-h). This charge transfer from Ag$_{20}$ to the substrate is also accompanied by notable charge redistribution within the cluster itself.

A significant amount of charge, 0.087 *e* per Ag atom, is transferred from Ag$_{20}$ to graphene (Figure 5e), associated with equilibration between the Fermi levels of the two systems. This leads to the strongest interfacial interaction. The amount of charge transfer between Ag$_{20}$ and the SiO$_2$ surface is larger for the rougher surfaces because of an increased local surface polarity that creates local electric fields. The smallest amount of charge is transferred from Ag$_{20}$ to smooth SiO$_2$, only 0.0013 *e* (Figure 5f), correlating with the smallest binding energy. The electron loss per Ag atom is 0.062 *e* and 0.058 *e* for the slightly and strongly rough SiO$_2$, also correlating with the binding energies. The simulations demonstrate that variation in the SiO$_2$ surface roughness can have a strong influence on adsorption of silver nanoparticles.

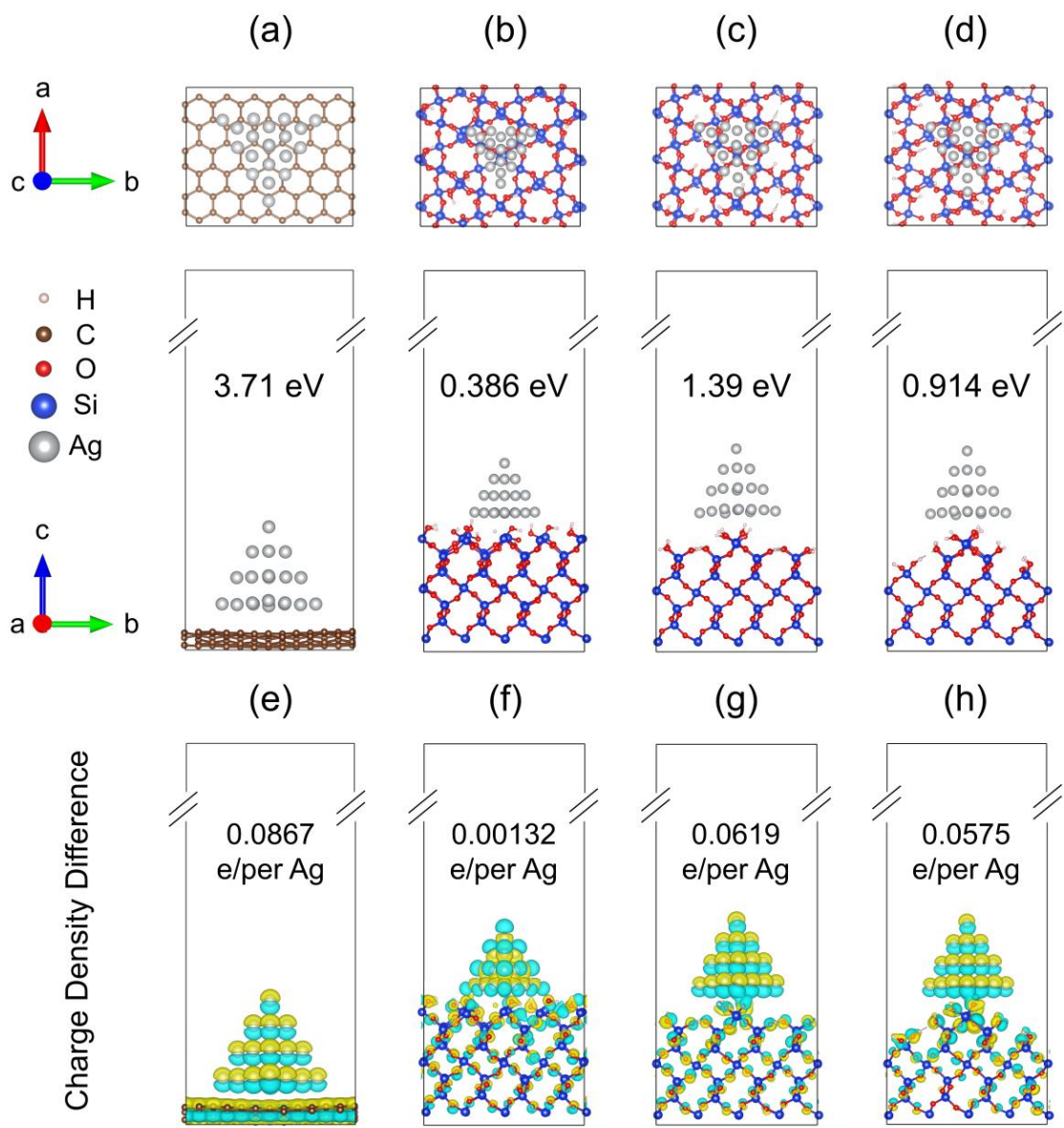

**Figure 5.** Top (top row) and side (middle row) views of the optimized geometries of **(a)** $Ag_{20}$/graphene, and $Ag_{20}$ interfaced with **(b)** smooth, **(c)** slightly rough, and **(d)** strongly rough $SiO_2$ surfaces. The binding energies are shown in the side-view panels. The charge density differences (isovalues: 0.06 a.u.) for the four investigated systems are displayed in **(e)-(h)**. Yellow: electron accumulation region. Blue: electron depletion region. The amount of charge transferred from $Ag_{20}$ to the substrate is shown in the panels.

In the present case, the post-treatment amorphous surface of $SiO_2$ was measured to display a root mean square surface roughness of ~0.2–0.3 nm. This places it near the category of a smooth surface. Furthermore, the nanoparticles used in this work are 6–10 nm in size, implying that adhesion forces are less influenced by charge transfer than those in the above model. Consequently, the experimentally observed contrast between graphene and silica surface coverage is concordant with the large difference between nanoparticle binding energies to these substrates, because weak binding enables the impinging nanoparticles to rebound upon surface impact.[49–51]

In conclusion, extremely high substrate selectivity was found for the coating of graphene devices by metal nanoparticles. Size-selective silver nanoparticle beam deposition onto properly precleaned devices decouples the coverage of the $SiO_2$ and graphene surfaces without the need for additional processing steps. This selectivity offers a significant advantage since it avoids the contamination and delamination risk posed by lithography processes, which is significant for graphene. It also avoids the surface modifications of noble metal nanoparticles required for deposition from colloidal suspensions. This observation may be of value for the fabrication both of single devices (cf. bismuth nanoparticle-assembled interconnects on patterned photoresist templates[52,53]) and of wafers containing many devices in close proximity. Graphene structures covered by noble metal thin films have attracted interest,[54–57] as have hybrid graphene-nanoparticle devices.

The data also highlight the high sensitivity of nanoparticle adsorption to surface contaminant removal. Model calculations of the interaction between silver nanoclusters and graphene or silica substrates confirm that particle binding to passivated $SiO_2$ surfaces is significantly weaker than that to graphene. The calculations also call attention to the role played by charge transfer from the nanoparticle to the substrate and by the roughness of the latter. In future work, it will be valuable to further explore the role of size effects in the interplay between particle charging energies, surface roughness, contact areas, and electron transfer phenomena. It also will be of interest to extend the work to additional substrate materials and nanoparticle sizes, shapes, and compositions, as well as velocities, where nonmonotonic behavior of adhesion probabilities have been predicted.[50]

**Supporting Information**

Supporting Information is available. Device fabrication details, images of three-monolayer coverage, deposition on graphene patches, and deposition on mica, description of contact-mode AFM manipulation of nanoparticles, and computational details.

**Notes**

The authors declare no competing financial interest

**Acknowledgments**

We thank M. Khojasteh, C. Niman, M. Chavez, M. El-Naggar, E. Levenson-Falk, and J. Chaney for help and discussions, and the staff of the Dornsife/Viterbi Machine Shop for expert technical assistance. This work was supported by the U.S. National Science Foundation under Grants DMR-2003469 (P.J.E., V.V.K.) and CHE-2154367 (O.V.P.). Work performed at The Aerospace Corporation was supported under The Aerospace Corporation's Independent Research and Development Program. R. L. acknowledges the National Science Foundation of China, Grant 21973006.

# REFERENCES

(1) Schedin, F.; Geim, A. K.; Morozov, S. V.; Hill, E. W.; Blake, P.; Katsnelson, M. I.; Novoselov, K. S. Detection of Individual Gas Molecules Adsorbed on Graphene. *Nat. Mater.* **2007**, *6*, 652–655.

(2) Varghese, S. S.; Lonkar, S.; Singh, K. K.; Swaminathan, S.; Abdala, A. Recent Advances in Graphene Based Gas Sensors. *Sensor. Actuat. B - Chem.* **2015**, *218*, 160–183.

(3) Han, W.; Kawakami, R. K.; Gmitra, M.; Fabian, J. Graphene Spintronics. *Nat. Nanotechnol.* **2014**, *9*, 794–807.

(4) Deng, X.; Wu, Y.; Dai, J.; Kang, D.; Zhang, D. Electronic Structure Tuning and Band Gap Opening of Graphene by Hole/Electron Codoping. *Phys. Lett. A* **2011**, *375*, 3890–3894.

(5) Amsterdam, S. H.; Marks, T. J.; Hersam, M. C. Leveraging Molecular Properties to Tailor Mixed-Dimensional Heterostructures beyond Energy Level Alignment. *J. Phys. Chem. Lett.* **2021**, *12*, 4543–4557.

(6) Abbas, G.; Sonia, F. J.; Jindra, M.; Červenka, J.; Kalbáč, M.; Frank, O.; Velický, M. Electrostatic Gating of Monolayer Graphene by Concentrated Aqueous Electrolytes. *J. Phys. Chem. Lett.* **2023**, *14*, 4281–4288.

(7) Scheerder, J. E.; Picot, T.; Reckinger, N.; Sneyder, T.; Zharinov, V. S.; Colomer, J.-F.; Janssens, E.; Van de Vondel, J. Decorating Graphene with Size-Selected Few-Atom Clusters: A Novel Approach to Investigate Graphene–Adparticle Interactions. *Nanoscale* **2017**, *9*, 10494–10501.

(8) Akbari-Sharbaf, A.; Ezugwu, S.; Ahmed, M. S.; Cottam, M. G.; Fanchini, G. Doping Graphene Thin Films with Metallic Nanoparticles: Experiment and Theory. *Carbon* **2015**, *95*, 199–207.

(9) Nguyen, K. T.; Li, D.; Borah, P.; Ma, X.; Liu, Z.; Zhu, L.; Grüner, G.; Xiong, Q.; Zhao, Y. Photoinduced Charge Transfer within Polyaniline-Encapsulated Quantum Dots Decorated on Graphene. *ACS Appl. Mater. Inter.* **2013**, *5*, 8105–8110.

(10) Dutta, R.; Pradhan, A.; Mondal, P.; Kakkar, S.; Sai, T. P.; Ghosh, A.; Basu, J. K. Enhancing Carrier Diffusion Length and Quantum Efficiency through Photoinduced Charge Transfer in Layered Graphene–Semiconducting Quantum Dot Devices. *ACS Appl. Mater. Inter.* **2021**, *13*, 24295–24303.

(11) Li, X.; Zhu, J.; Wei, B. Hybrid Nanostructures of Metal/Two-Dimensional Nanomaterials for Plasmon-Enhanced Applications. *Chem. Soc. Rev.* **2016**, *45*, 3145–3187.

(12) Grigorenko, A. N.; Polini, M.; Novoselov, K. Graphene Plasmonics. *Nat. Photonics* **2012**, *6*, 749–758.

(13) Cherqui, C.; Li, G.; Busche, J. A.; Quillin, S. C.; Camden, J. P.; Masiello, D. J. Multipolar Nanocube Plasmon Mode-Mixing in Finite Substrates. *J. Phys. Chem. Lett.* **2018**, *9*, 504–512.

(14) Wang, T.; Huang, D.; Yang, Z.; Xu, S.; He, G.; Li, X.; Hu, N.; Yin, G.; He, D.; Zhang, L. A Review on Graphene-Based Gas/Vapor Sensors with Unique Properties and Potential Applications. *Nano-Micro Lett.* **2016**, *8*, 95–119.

(15) Ruffino, F.; Giannazzo, F. A Review on Metal Nanoparticles Nucleation and Growth on/in Graphene. *Crystals* **2017**, *7*, 219.

(16) Shtepliuk, I.; Yakimova, R. Computational Appraisal of Silver Nanocluster Evolution on Epitaxial Graphene: Implications for CO Sensing. *ACS Omega* **2021**, *6*, 24739–24751.

(17) Cui, S.; Mao, S.; Lu, G.; Chen, J. Graphene Coupled with Nanocrystals: Opportunities and Challenges for Energy and Sensing Applications. *J. Phys. Chem. Lett.* **2013**, *4*, 2441–2454.
(18) Kessler, B.; Girit, Ç.; Zettl, A.; Bouchiat, V. Tunable Superconducting Phase Transition in Metal-Decorated Graphene Sheets. *Phys. Rev. Lett.* **2010**, *104*, 047001.
(19) Allain, A.; Han, Z.; Bouchiat, V. Electrical Control of the Superconducting-to-Insulating Transition in Graphene–Metal Hybrids. *Nat. Mater.* **2012**, *11*, 590–594.
(20) Han, Z.; Allain, A.; Arjmandi-Tash, H.; Tikhonov, K.; Feigel'Man, M.; Sacépé, B.; Bouchiat, V. Collapse of Superconductivity in a Hybrid Tin–Graphene Josephson Junction Array. *Nat. Phys.* **2014**, *10*, 380–386.
(21) Milani, P.; Iannotta, S. *Cluster Beam Synthesis of Nanostructured Materials*; Springer: Berlin, 1999.
(22) *Gas-Phase Synthesis of Nanoparticles*; Huttel, Y., Ed.; Wiley-VCH: Weinheim, 2017.
(23) *Cluster Beam Deposition of Functional Nanomaterials and Devices*; Milani, P.; Sowwan, M., Eds.; Elsevier: Amsterdam, 2020.
(24) Her, M.; Beams, R.; Novotny, L. Graphene Transfer with Reduced Residue. *Phys. Lett. A* **2013**, *377*, 1455–1458.
(25) Xie, W.; Weng, L.-T.; Ng, K. M.; Chan, C. K.; Chan, C.-M. Clean Graphene Surface through High Temperature Annealing. *Carbon* **2015**, *94*, 740–748.
(26) Yu, Q.; Lian, J.; Siriponglert, S.; Li, H.; Chen, Y. P.; Pei, S.-S. Graphene Segregated on Ni Surfaces and Transferred to Insulators. *Appl. Phys. Lett.* **2008**, *93*, 113103.
(27) Reina, A.; Jia, X.; Ho, J.; Nezich, D.; Son, H.; Bulovic, V.; Dresselhaus, M. S.; Kong, J. Large Area, Few-Layer Graphene Films on Arbitrary Substrates by Chemical Vapor Deposition. *Nano Lett.* **2009**, *9*, 30–35.
(28) Li, X.; Cai, W.; An, J.; Kim, S.; Nah, J.; Yang, D.; Piner, R.; Velamakanni, A.; Jung, I.; Tutuc, E. Large-Area Synthesis of High-Quality and Uniform Graphene Films on Copper Foils. *Science* **2009**, *324*, 1312–1314.
(29) Haberland, H.; Mall, M.; Moseler, M.; Qiang, Y.; Reiners, T.; Thurner, Y. Filling of Micron-sized Contact Holes with Copper by Energetic Cluster Impact. *J. Vac. Sci. Technol. A* **1994**, *12*, 2925–2930.
(30) Khojasteh, M. Fabrication, Deposition, and Characterization of Size-Selected Metal Nanoclusters With a Magnetron Sputtering Gas Aggregation Source, University of Southern California, 2019.
(31) Johnson, G. E.; Colby, R.; Laskin, J. Soft Landing of Bare Nanoparticles with Controlled Size, Composition, and Morphology. *Nanoscale* **2015**, *7*, 3491–3503.
(32) San Paulo, A.; García, R. High-Resolution Imaging of Antibodies by Tapping-Mode Atomic Force Microscopy: Attractive and Repulsive Tip-Sample Interaction Regimes. *Biophys. J.* **2000**, *78*, 1599–1605.
(33) Khojasteh, M.; Kresin, V. V. Influence of Source Parameters on the Growth of Metal Nanoparticles by Sputter-Gas-Aggregation. *Appl. Nanosci.* **2017**, *7*, 875–883.
(34) Francis, G.; Kuipers, L.; Cleaver, J.; Palmer, R. Diffusion Controlled Growth of Metallic Nanoclusters at Selected Surface Sites. *J. Appl. Phys.* **1996**, *79*, 2942–2947.
(35) Carroll, S.; Seeger, K.; Palmer, R. Trapping of Size-Selected Ag Clusters at Surface Steps. *Appl. Phys. Lett.* **1998**, *72*, 305–307.
(36) Nicolet, M.-A. Diffusion Barriers in Thin Films. *Thin Solid Films* **1978**, *52*, 415–443.

(37) Khanna, V. Adhesion–Delamination Phenomena at the Surfaces and Interfaces in Microelectronics and MEMS Structures and Packaged Devices. *J. Phys. D Appl. Phys.* **2011**, *44*, 034004.

(38) Romanyuk, A.; Steiner, R.; Mack, I.; Oelhafen, P.; Mathys, D. Growth of Thin Silver Films on Silicon Oxide Pretreated by Low Temperature Argon Plasma. *Surf. Sci.* **2007**, *601*, 1026–1030.

(39) Aono, T.; Iwasaki, T.; Yoshimura, Y.; Nakayama, Y. Method for Fabricating Multilevel Interconnection Structures Using Differential Delamination Energies between Metal and Silicon Oxide Thin Films. *Electr. Commun. Jpn.* **2016**, *99*, 41–49.

(40) Nakada, K.; Ishii, A. Migration of Adatom Adsorption on Graphene Using DFT Calculation. *Solid State Commun.* **2011**, *151*, 13–16.

(41) Sorescu, D. C.; Thompson, D. L.; Hurley, M. M.; Chabalowski, C. F. First-Principles Calculations of the Adsorption, Diffusion, and Dissociation of a CO Molecule on the Fe (100) Surface. *Phys. Rev. B* **2002**, *66*, 035416.

(42) Chu, W.; Saidi, W. A.; Prezhdo, O. V. Long-Lived Hot Electron in a Metallic Particle for Plasmonics and Catalysis: *Ab Initio* Nonadiabatic Molecular Dynamics with Machine Learning. *ACS Nano* **2020**, *14*, 10608–10615.

(43) Rapacioli, M.; Spiegelman, F.; Tarrat, N. Evidencing the Relationship between Isomer Spectra and Melting: The 20-and 55-Atom Silver and Gold Cluster Cases. *Phys. Chem. Chem. Phys.* **2019**, *21*, 24857–24866.

(44) Ren, X.; Lin, W.; Fang, Y.; Ma, F.; Wang, J. Raman Optical Activity (ROA) and Surface-Enhanced ROA (SE-ROA) of (+)-(R)-Methyloxirane Adsorbed on a $Ag_{20}$ Cluster. *RSC Adv.* **2017**, *7*, 34376–34381.

(45) Liu, L.; Chen, D.; Ma, H.; Liang, W. Spectral Characteristics of Chemical Enhancement on SERS of Benzene-like Derivatives: Franck–Condon and Herzberg–Teller Contributions. *J. Phys. Chem. C* **2015**, *119*, 27609–27619.

(46) Gao, W.; Xiao, P.; Henkelman, G.; Liechti, K. M.; Huang, R. Interfacial Adhesion between Graphene and Silicon Dioxide by Density Functional Theory with van Der Waals Corrections. *J. Phys. D Appl. Phys.* **2014**, *47*, 255301.

(47) Fan, X.; Zheng, W.; Chihaia, V.; Shen, Z.; Kuo, J.-L. Interaction Between Graphene and the Surface of $SiO_2$. *J. Phys. - Condens. Mat.* **2012**, *24*, 305004.

(48) Henkelman, G.; Arnaldsson, A.; Jónsson, H. A Fast and Robust Algorithm for Bader Decomposition of Charge Density. *Comp. Mater. Sci.* **2006**, *36*, 354–360.

(49) Awasthi, A.; Hendy, S.; Zoontjens, P.; Brown, S. Reentrant Adhesion Behavior in Nanocluster Deposition. *Phys. Rev. Lett.* **2006**, *97*, 186103.

(50) Awasthi, A.; Hendy, S.; Zoontjens, P.; Brown, S.; Natali, F. Molecular Dynamics Simulations of Reflection and Adhesion Behavior in Lennard-Jones Cluster Deposition. *Phys. Rev. B* **2007**, *76*, 115437.

(51) Weir, G. A Simple Conceptual Model for the Behaviour of an Impacting Rigid-Plastic, Spherical, Nano-Scale Particle. *Curr. Appl. Phys.* **2008**, *8*, 355–358.

(52) Reichel, R.; Partridge, J.; Natali, F.; Matthewson, T.; Brown, S.; Lassesson, A.; Mackenzie, D.; Ayesh, A.; Tee, K.; Awasthi, A. From the Adhesion of Atomic Clusters to the Fabrication of Nanodevices. *Appl. Phys. Lett.* **2006**, *89*, 213105.

(53) Partridge, J.; Matthewson, T.; Brown, S. Bi Cluster-Assembled Interconnects Produced Using SU8 Templates. *Nanotechnology* **2007**, *18*, 155607.

(54)  Rast, L.; Sullivan, T.; Tewary, V. K. Stratified Graphene/Noble Metal Systems for Low-Loss Plasmonics Applications. *Phys. Rev. B* **2013**, *87*, 045428.
(55)  Yakubovsky, D. I.; Stebunov, Y. V.; Kirtaev, R. V.; Voronin, K. V.; Voronov, A. A.; Arsenin, A. V.; Volkov, V. S. Graphene-Supported Thin Metal Films for Nanophotonics and Optoelectronics. *Nanomaterials* **2018**, *8*, 1058.
(56)  Shtepliuk, I.; Ivanov, I. G.; Pliatsikas, N.; Iakimov, T.; Jamnig, A.; Sarakinos, K.; Yakimova, R. Probing the Uniformity of Silver-Doped Epitaxial Graphene by Micro-Raman Mapping. *Physica B* **2020**, *580*, 411751.
(57)  Chahal, S.; Bandyopadhyay, A.; Dash, S. P.; Kumar, P. Microwave Synthesized 2D Gold and Its 2D-2D Hybrids. *J. Phys. Chem. Lett.* **2022**, *13*, 6487–6495.

# Substrate-Selective Adhesion of Metal Nanoparticles to Graphene Devices

Patrick J. Edwards[1,2], Sean C. Stuart[2], James T. Farmer[1], Ran Shi,[3] Run Long,[3] Oleg V. Prezhdo[1,4], Vitaly V. Kresin[1]

[1] Department of Physics and Astronomy, University of Southern California, Los Angeles, CA 90089-0484, USA

[2] Physical Sciences Laboratories, The Aerospace Corporation, 355 S. Douglas St., El Segundo, CA 90245, USA

[3] College of Chemistry, Key Laboratory of Theoretical and Computational Photochemistry of Ministry of Education, Beijing Normal University, Beijing 100875, China

[4] Department of Chemistry, University of Southern California, Los Angeles, CA 90089, USA

Contents

## S-I. Device fabrication

Graphene FET devices were fabricated in a cleanroom environment at the Aerospace Corporation. Square 1 cm × 1 cm dies were cut from a wafer of CVD graphene grown on a Si/SiO$_2$ substrate. Each die was then patterned with a 5 × 5 array of four-probe graphene FETs via a multi-step lithography process. The rectangular graphene channel was defined by electron beam lithography (EBL) and etched out of the graphene layer by using an argon plasma. Then an additional EBL and electron beam physical vapor deposition (EBPVD) process was performed to deposit four Ti/Au (10nm/50 nm) contacts onto the channel surface, completing the device. An image of one FET is shown in Figure S1.

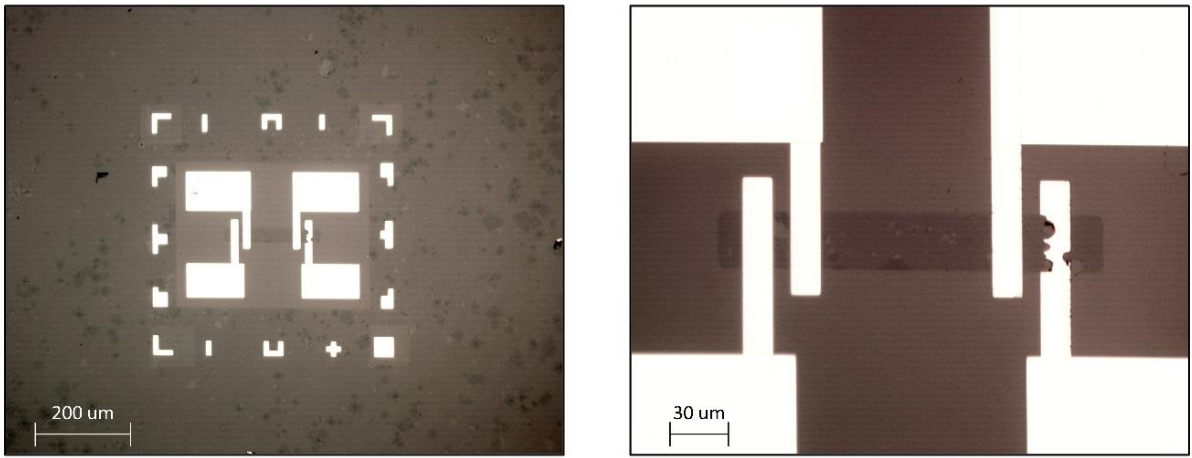

**Figure S1**. Optical microscopy imaging of a completed graphene device showing an isolated graphene channel (dark strip in the center of the image) on an SiO$_2$ substrate with Ti/Au surface contacts.

## S-II. Three-monolayer nanoparticle coverage

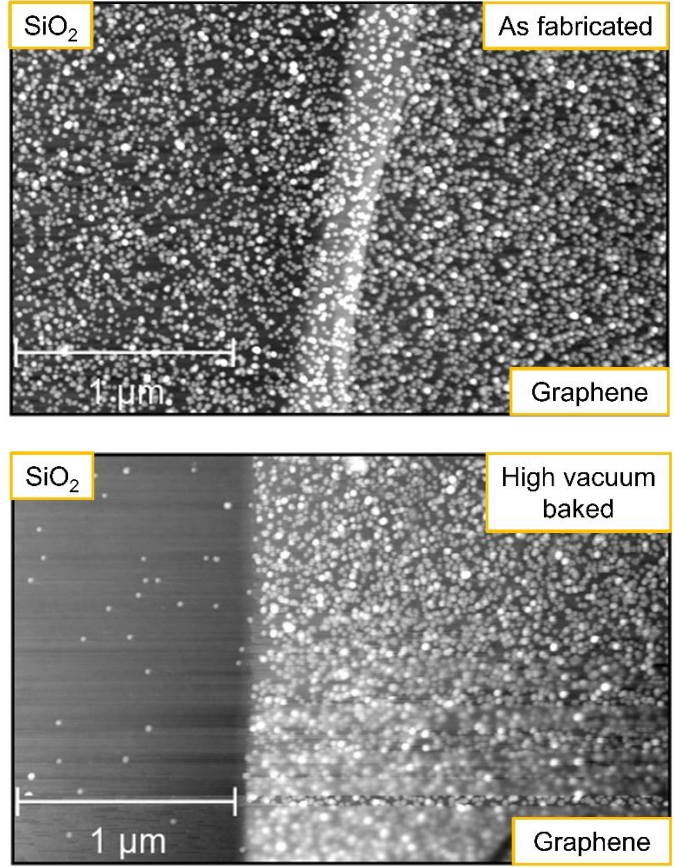

**Figure S2.** AFM images of graphene FET devices after the deposition of three monolayers of Ag nanoparticles. The effect of residue removal pre-treatment by device baking on the nanoparticle adhesion is clearly visible. (Loss of resolution at the bottom of the lower image is due to partial tip disengagement during scanning.)

## S-III. Nanoparticle deposition on graphene patches

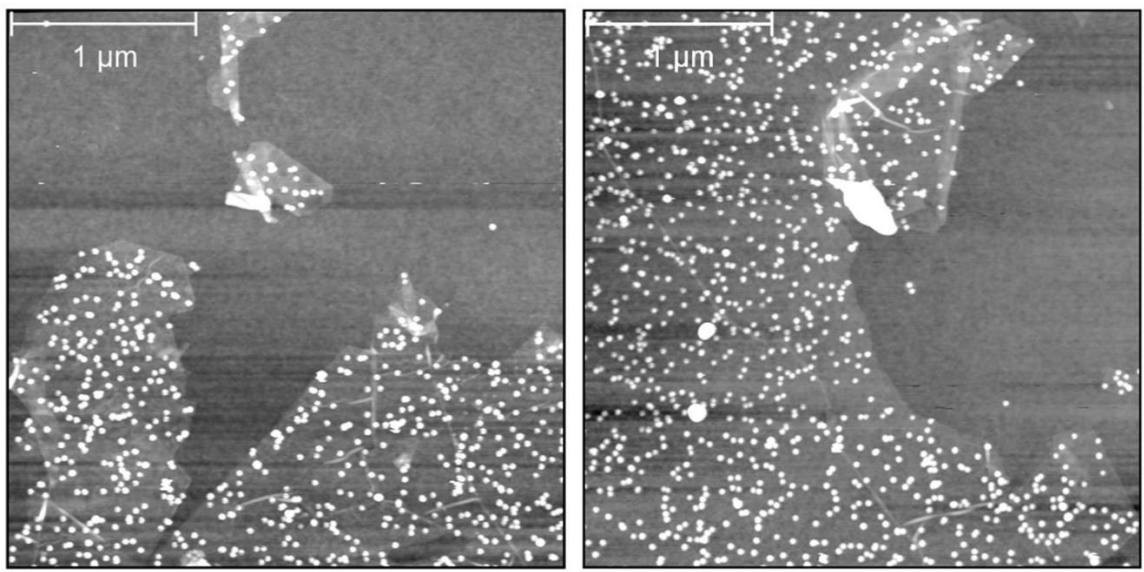

**Figure S3.** Silver nanoparticles cover electrically isolated islands of graphene but not the surrounding expanse of $SiO_2$. There also is no evidence of nanoparticle aggregation along the graphene step edges, and hence of any significant surface diffusion.

## S-IV. Nanoparticle deposition on mica

A disc of mica (Ted Pella, optically flat grade V1) was cleaved in air and mounted adjacent to a wafer of CVD graphene in the same manner as the samples described in the main text. Analogously to the graphene substrate, the mica surface possesses an organized crystal structure and provides a very flat surface upon which to deposit nanoparticles.[S1, S2] However, analogously to silica, it is electrically insulating. Imaging shows that the mica and the graphene substrates become similarly covered.

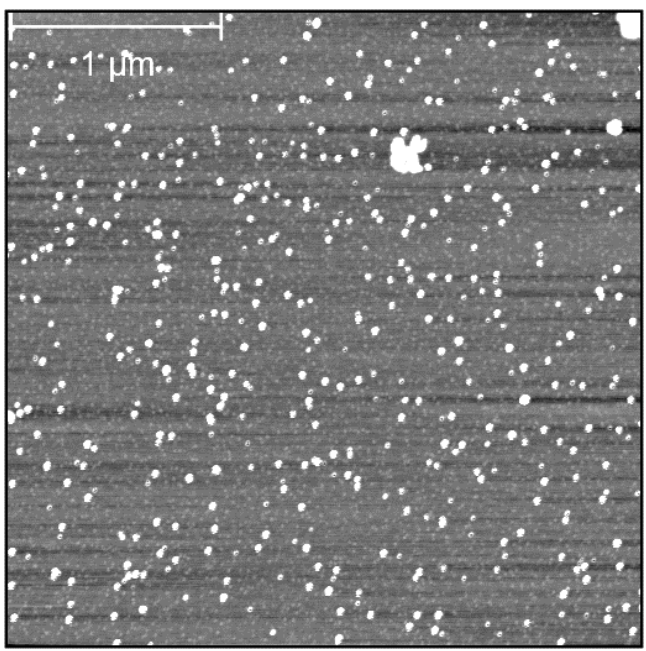

Figure **S4.** Insulating mica samples show a similar degree of nanoparticle coverage as the adjacent CVD graphene substrates.

## S-V. Contact-mode nanomanipulation

Nanomanipulation experiments were conducted using Ir/Ti coated conductive silicon tips (Oxford Instruments ASYELEC.01-R2). The procedure is first to obtain a small (~1 μm$^2$) field-of-view image of the deposited nanoparticles near a graphene/SiO$_2$ step edge (see Figure S4 top left). This first image is acquired in the non-contact attractive regime, so as not to disturb the particles, as described in the main text. Once this image is obtained, the system allows for seamless transition to contact-mode imaging where custom tip deflection settings and paths can be preset by using the collected image as a reference (see the path overlay in Figure S4 top right). The paths are then traced out, in order, while the tip is in contact with the sample surface. For all contact traces the tip speed is set to the slowest possible scan rate of 5 nm/s, and the tip lifts off the surface when relocating from the end of one trace to the beginning of the next. After all contact traces are completed, the image area is finally rescanned in the non-contact mode to reveal the result of the manipulations. Figure S4 depicts a representative manipulation cycle and displays the range of observed outcomes.

Trace 1 shows the move of a nanoparticle off the graphene onto the cleaned SiO$_2$ surface.

Trace 2 demonstrates a frequent occurrence for a nanoparticle that had already been moved to the SiO$_2$ during a previous manipulation. This trace sought simply to move it along the oxide surface, but the post-manipulation image finds that the particle has been "erased" from the region (Figure S4 bottom). What happened is that the nanoparticle desorbed from the substrate and attached itself to the Ir/Ti coated tip. Such an outcome also has been observed while attempting to move nanoparticles from graphene to SiO$_2$ (akin to trace 1).

Finally, trace 3 relocated a cluster along the graphene toward the SiO$_2$ interface. A close inspection of the result reveals not one but two nanoparticles at the final location (highlighted by the blue box). Thus, it appears that the above tip-adsorbed nanoparticle has been released onto the graphene surface.

The attachment of nanoparticles to the tip was observed only when moving them along the clean SiO$_2$ substrate, and nanoparticle were never removed from the graphene. Additionally, nanoparticle redeposition was only observed onto the graphene surface. This suggests that while the nanoparticle interaction with the tip is stronger than that with the oxide surface, the nanoparticle-graphene interaction is stronger still.

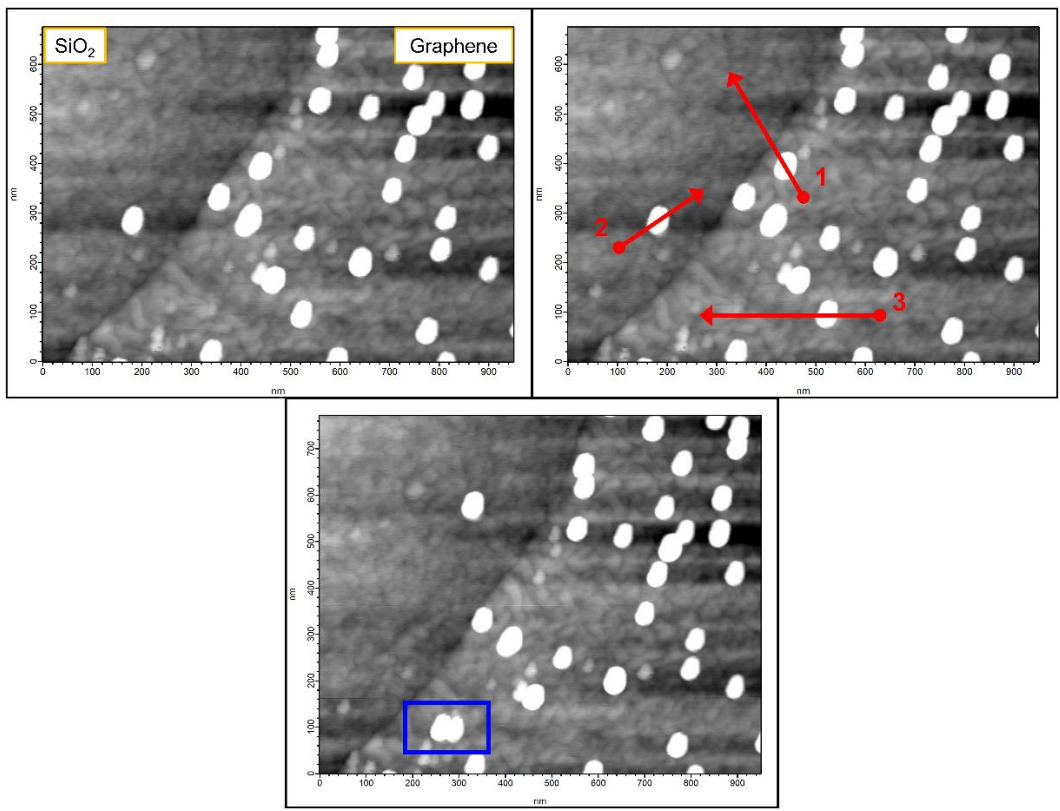

**Figure S5.** Top left: initial image of nanoparticles deposited near a graphene-$SiO_2$ step edge. Top right: the same image overlayed with the defined contact mode traces showing the programmed particle manipulation paths along the surface. Bottom: image of the outcome of the traced paths.

## S-VI. Computational details

The calculations are carried out within the density functional theory (DFT) framework using the Vienna *ab initio* simulation package (VASP)[S3] which employs periodic boundary conditions and plane-wave basis sets. The projector augmented wave (PAW)[S4] method is used to describe the electron-ion interactions, and the Perdew-Burke-Ernzerhof (PBE)[S5] functional is applied to account for the electronic exchange-correlation interactions. A uniform 1×1×1 Monkhorst-Pack *k*-point sampling[S6] and a plane-wave energy cutoff of 400 eV are utilized. The van der Waals interactions are described using the Grimme DFT-D3 method with the Becke-Johnson damping.[S7] The geometry optimization is considered converged when ion forces become less than $10^{-3}$ eV·Å$^{-1}$. The SiO$_2$ surface contains silanol groups (Si-OH) forming a zigzag hydrogen bonded network. A 20 Å vacuum layer is added to the surface normal in all systems to avoid spurious interactions between the periodic images.

## S-VII. References

(S1) Senden, T. J.; Ducker, W. A. Surface Roughness of Plasma-Treated Mica. *Langmuir* **1992**, *8*, 733-735.

(S2) Ostendorf, F.; Schmitz, C.; Hirth, S.; Kühnle, A.; Kolodziej, J. J.; Reichling, M. How Flat is an Air-Cleaved Mica Surface? *Nanotechnology* **2008**, *19*, 305705-305705.

(S3) Kresse, G.; Furthmüller, J. Efficient Iterative Schemes for *Ab Initio* Total-Energy Calculations Using a Plane-Wave Basis Set. *Phys. Rev. B* **1996**, *54*, 11169-11186.

(S4) Blöchl, P. E. Projector Augmented-Wave Method. *Phys. Rev. B* **1994**, *50*, 17953-17979.

(S5) Perdew, J. P.; Burke, K.; Ernzerhof, M. Generalized Gradient Approximation Made Simple. *Phys. Rev. Lett.* **1996**, *77*, 3865-3868.

(S6) Monkhorst, H. J.; Pack, J. D. Special Points for Brillouin-Zone Integrations. *Phys. Rev. B* **1976**, *13*, 5188-5192.

(S7) Grimme, S.; Antony, J.; Ehrlich, S.; Krieg, H. A Consistent and Accurate *Ab Initio* Parametrization of Density Functional Dispersion Correction (DFT-D) for the 94 Elements H-Pu. *J. Chem. Phys.* **2010**, *132*.